# Flexible orbital torque device with ultralow switching current

Liguang Gong,[1,2,a)] Jian Song,[1,3,a)] Bin Lao,[1] Run-Wei Li,[1,4,b)] Zhiming Wang[1,2,b)]

[1]Ningbo Institute of Materials Technology and Engineering, Chinese Academy of Sciences, Ningbo 315201, China.

[2]Center of Materials Science and Optoelectronics Engineering, University of Chinese Academy of Sciences, Beijing 100049, China.

[3]Nano Science and Technology Institute, University of Science and Technology of China, Suzhou, China.

[4]Eastern Institute of Technology, Ningbo 315200, China

[a)]Authors who contributed equally to this work

[b)]Authors to whom correspondence should be addressed: runweili@nimte.ac.cn; zhiming.wang@nimte.ac.cn

**ABSTRACT**

Orbital torque (OT) offers a highly efficient way for electrical magnetization manipulation. However, its potential in the emerging field of flexible spintronics remains largely unexplored. Here, we demonstrate a flexible and robust OT device based on a mica/SrRuO$_3$(SRO)/CoPt heterostructure. We measure a large torque efficiency of -0.31, which originates from the significant orbital Hall effect in the SRO layer. Leveraging the low thermal conductivity of the mica substrate, a thermally-assisted switching mechanism is activated, enabling an ultralow threshold current density of $9.2 \times 10^9$ A/m$^2$. This value represents a 90% reduction compared to conventional spin-torque devices and a 52% reduction against its rigid counterpart on a SrTiO$_3$ substrate. The superior performances is well-maintained after $10^3$ bending cycles, conforming its exceptional flexibility and durability. Our work pioneers the development of flexible OT devices, showcasing a viable path toward next-generation, low-power wearable spintronic applications.

Flexible spintronic devices, which integrate the portability of flexible electronics with the non-volatility, high speed of spintronics,[1-4] hold significant promise for next-generation wearable information technologies.[5-10] A pivotal element for their practical application is the efficient electrical manipulation of magnetization, typically achieved via spin orbit torque (SOT).[11-15] To date, flexible SOT devices have predominantly relied on heavy metals like platinum deposited on organic substrates (e.g. PI, PEN).[9,10,16-21] Although the low thermal conductivity of these substrates introduces a beneficial heat-assist effect,[17,18,22,23] the intrinsic inefficiency of the spin Hall effect in these materials means that the switching current density remians prohibitively high, typically around $10^{11}$ A/m².[9,10,16-19] This fundamental limitation severely hampers the development of low-power wearable electronics.

The recently discovered orbital torque (OT), originates from orbital Hall effect (OHE), offers a highly competitive alternative for efficient magnetization control.[24-26] Unlike the spin Hall effect, which relies heavily on strong spin-orbit coupling in heavy metals, the OHE arises from the intrinsic orbital texture of a material's band structure, promising higher torque efficiencies even in lighter materials.[27-32] Furthermore, the OHE exhibits a unique and complex response to extrinsic scattering, potentially leading to a paradoxical enhancement of orbital current generation.[33-37] Recent work on transition metal oxide $SrRuO_3$ (SRO) has unveiled an unconventional scaling law of the OHE, where disorder boosts orbital Hall conductivity.[38] This scattering-enhanced torque mechanism, a paradigm unavailable in conventional spin-torque systems, enabled a substantial reduction of both switching current density and power consumption on rigid substrates. This discovery presents a compelling opportunity of OT-based approaches for low-power flexible spintronic devices. However, the integration of OT onto flexible platforms and their actual magnetization switching performance remain large unexplored.

Here, we demonstrate a flexible OT device based on a single-crystalline SRO/CoPt heterostructure on a mica substrate. The device exhibits a significant torque efficiency of -0.31, stemming from the prominent OHE in SRO. Crucially, we unveil a synergistic interplay between the efficient OT and a thermally-assisted switching mechanism, the latter enabled by the low thermal conductivity of the mica substrate. This synergy leads to an exceptionally low threshold current density of $9.2 \times 10^9$ A/m², representing a ~90% reduction compared to flexible spin torque devices and a 52% reduction against its rigid counterpart. Furthermore, the device exhibits outstanding mechanical robustness, with its superior performances maintained after $10^3$ bending cycles. Our work pioneers the practical potential of OT in flexible eletronics, showcase a viable path toward next-generation energy-efficient wearable spintronics.

The thin flims with the structure of SrTiO$_3$(15)/SrRuO$_3$(20)/[Pt(1)Co(0.5)]$_2$Pt(1) (numbers in parentheses denote thickness in nanometers), as shown in Fig. 1(a), were fabricated on mica substrates via pulsed laser deposition (PLD) and magnetron sputtering. Detailed sample preparation and material characterization data are provided in Supplementary Material S1. Here, SrTiO$_3$(STO) serves as a buffer layer to achieve high-quality (111)-oriented SrRuO$_3$(SRO) on the mica substrate. Owing to the coexistence of spin Hall effect (SHE) and orbital Hall effect (OHE) in SRO,[38] the applied current can be converted into both spin and orbital currents, which are subsequently injected into the CoPt multilayer. The CoPt system not only exhibits excellent perpendicular magnetic anisotropy (PMA) but also possesses prominent orbital-to-spin conversion coefficients ($\eta_{L\text{-}S}$),[39] enabling efficient conversion of orbital currents into spin currents. However, the spins generated through these two pathways exhibit opposite orientations, consequently resulting in opposite signs between the spin torque and orbital torque.

Following standard photolithography and ion milling techniques, the film stacks were patterned into Hall cross devices (70 μm in length and 15 μm in width). We performed the harmonic Hall voltage (HHV) measurements on the Hall cross devices to quantify the spin orbit torque (SOT) efficiency.[40] An alternating current $I_{ac}$ with the frequency of 133 Hz was applied, the fist ($R_{xy}^{1\omega}$) and the second harmonic Hall resistance ($R_{xy}^{2\omega}$) were measured as functions of the magnetic field applied along the current direction ($H_x$), as shown in Fig. 1(b) and 1(c). For the perpendicular magnetic anisotropy (PMA) system, $R_{xy}^{1\omega}$ is related to the rotation of magnetic moments under in-plane magnetic fields. The magnetic anisotropy field $H_k$ can be determined to be 3330 Oe through fitting using the relation $R_{xy}^{1\omega} = R_H\sqrt{1-(H_x/H_k)^2}$, where $R_H$ represents anomalous Hall resistance (the amplitude of $R_{xy}^{1\omega}$). The second harmonic Hall resistance $R_{xy}^{2\omega}$ arises from oscillations of the net magnetic moment under the torque-induced effective field (dominated by the damping-like effective field $H_{DL}$ in SRO).[41-44] When the in-plane magnetic field $H_x$ exceeds $H_k$, $R_{xy}^{2\omega}$ can be written as

$$R_{xy}^{2\omega} = \text{sgn}(H_x)\left(\frac{R_H}{2}\frac{H_{DL}}{|H_x|-H_k} + R_{\text{offset}}\right) \quad (1)$$

Where $R_{\text{offset}}$ is mainly contributed by anomalous Nernst effect and misalignment of the Hall cross.[45,46] $H_{DL}$ was calculated by fitting the slop of $R_{xy}^{2\omega}$ vs. $1/(|H_x|-H_k)$ under various currents in the Fig. 1(d), with the results summarized as hollow dots in the Fig. 1(e). Notably, as the current density $J$ decreases, $H_{DL}$ exhibits negative values and decreases linearly, consistent with the previously reported orbital torque-dominated origin in SRO systems. The torque efficiency $\xi_{DL}$ was then calculated using the relation $\xi_{DL} = 2eM_st_{FM}H_{DL}/\hbar J$, where e is the electron charge, $\hbar$ is the reduced Planck constant, $M_s$ and $t_{FM}$ are the saturation magnetization and thickness of the CoPt layer. With the measured $M_s$ of 490 emu/cm$^3$ (see the supplementary material, Fig. S1 (c)), the $\xi_{DL}$ for SRO was determined to be -0.31. The negative sign indicates that the negative orbital Hall conductivity dominates over positive spin Hall conductivity in SRO.[38] The high absolute value demonstrates that SRO(111) on our mica substrate maintains the advantage of torque efficiency previously observed on the rigid STO(001) substrate.

To further investigate the practical efficiency of the magnetization switching of SRO/CoPt system on mica substrates, we conducted current-induced magnetization switching experiments at room temperature on both mica substrates and control STO(111) substrates. Figures 2(a) and 2(b) show the anomalous Hall resistance ($R_{xy}$) as a function of out-of-plane magnetic field ($H_z$) for mica/SRO/CoPt and STO(111)/SRO/CoPt samples, respectively. The square-shaped anomalous Hall loops confirm robust PMA in both systems. Subsequently, a small dc current of 100 µA was applied to read the Hall signal, while multiple write pulse current with alternative magnitudes and width of 200 µs were injected. An in-plane external magnetic field was applied along the current direction to break the system's symmetry. As shown in Fig. 2(c), for the device on mica substrate under a 100 Oe in-plane external field, the Hall resistance begins to change when the write current density increases from 0 to $9.2 \times 10^9$ A/m$^2$, indicating the beginning of magnetization switching. This current density is defined as the threshold current density $J_{th}$. When the external field reverses from +100 Oe to -100 Oe, the switching loop changes from counterclockwise to clockwise, and disappears entirely at zero field, confirming deterministic magnetization switching. For the control sample fabricated on STO substrate (Fig. 2(d)), identical magnetization switching behavior was observe, While $J_{th}$ increases to $2.3 \times 10^{10}$ A/m$^2$. The two type substures were tested with multiple devices, showing insignificant variations in $J_{th}$ for each substure, as represented by error bars in Fig. 2(e). This demonstrates that the reduction in $J_{th}$ for mica-based samples compared to STO substrates is highly stable.

For reference, Fig. 2(e) also compiles $J_{th}$ values of state-of-the-art flexible Pt-based spin torque devices.[9,10,16-19] It is noteworthy that nearly all previous studies on flexible spin torque devices have utilized direct deposition of heavy metal Pt layers on polymer substrates (e.g., PI, PEN) to generate spin currents. Excitingly, our flexible mica/SRO/CoPt devices achieves a $J_{th}$ of $9.2 \times 10^9$ A/m$^2$, indicates a reduction by approximately 90% compared to the average $J_{th}$ ($9.1 \times 10^{10}$ A/m$^2$) observed in flexible Pt-based devices. Considering the resistivity ($\rho$) difference between Pt ($\approx 50$ μΩ·cm) [20,21] and SRO (400 μΩ·cm in this work), the power consumption ($\propto J_{th}^2 \cdot \rho$) of our flexible device represents only 8% of the average power consumption level in flexible Pt-based systems. This remarkable improvement in power efficiency can be attributed to two factors, as presented in the comparative discussion below. First, both rigid and flexible SRO-based devices demonstrate a lower $J_{th}$ than Pt-based counterparts, suggesting that the intrinsically efficient OT of SRO itself plays a crucial role. Second, analogous to the heat-assisted effects induced current density reduction observed when transitioning from rigid to flexible Pt devices,[18] a 52% reduction of $J_{th}$ occurs in SRO devices on mica versus STO substrates. Given the comparable torque efficiencies observed for SRO/CoPt bilayers on both mica and STO substrates (-0.29 for STO-based devices, Supplementary Fig. S2), coupled with the intrinsically low thermal conductivity of mica (4.05 W·m$^{-1}$·K$^{-1}$, ref[47]), this observation strongly suggests the presence of mica substrate-induced heat-assisted effects that further reduces the switching current in the mica/SRO/CoPt devices.

Next, we quantitatively investigated the heating capability of devices on mica substrates and their contribution to assisted magnetization switching. The temperature-dependent and the pulse current-dependent longitudinal resistance change ($\Delta R_{xx}$) were measured on identical Hall cross devices fabricated on both mica and STO substrates, as shown in Fig. S3 and Fig. 3(a). It can be observed that the resistance response to temperature is similar for both substrates. However, the pulsed current induces a significantly faster rise in resistance on the mica substrate compared to the STO substrate. Using equivalent resistance changes as a bridge, the relationship between the applied pulsed current and the sample temperature was established for both substrates, as presented in Fig. 3(b). The dashed lines indicate the threshold switching currents and corresponding sample temperatures for magnetization reversal on each substrate (detailed magnetization switch of these two samples are shown in supplementary material S4). For the mica substrate, magnetization switching begins at a threshold current of 10 mA, corresponding to a sample temperature of 437 K. In contrast, on the STO substrate, the threshold current of 24 mA corresponds to a sample temperature of 425 K. Clearly, elevating the sample temperature from room temperature to approximately 430 K requires a significantly lower pulsed current on the mica substrate. This indicates that the lower thermal conductivity of mica suppresses the dissipation of Joule heat generated by the pulsed current, resulting in a higher localized sample temperature during current pulse application.

To investigate the contribution of significant temperature rise in the mica substrate to the magnetization switching process, we compared the reduction in anomalous Hall resistance $R_H$ (proportional to the saturation magnetization $M_s$) and magnetic anisotropy field $H_k$ related to the switching barrier for CoPt magnetic layers on mica and STO substrates. As shown in Fig. 3(c), as the pulsed current applied to the mica-based sample increased from 0.1 mA to 13 mA, $R_H$ remained almost constant until the current exceeded 10 mA, after which it decreased significantly with further current increase. While the phenomenon on the STO substrate was similar, $R_H$ exhibited a pronounced decrease only above a significantly higher threshold current of 24 mA (Fig. 3(d)). Despite this difference in threshold current, Fig. 3(b) demonstrates that they correspond to the same critical temperature. This equivalence is further demonstrated by heating Hall measurements (Fig. 3(e) and Fig. S4), where $R_H$ significantly decreases only above 400 K for both samples. Using the established relationship between pulsed current and temperature, we compiled the $R_H$ data obtained from both pulse and heating Hall tests for both substrates into Fig. 3(f). The resulting curves demonstrate remarkable agreement. Furthermore, the pulsed current and temperature dependence of $H_k$ extracted from in-plane Hall measurements ($R_{xy} - H_x$, Supplementary Fig. S5) exhibited behavior consistent with that of $R_H$. These results collectively demonstrate that Joule heating contributes to the reduction of both magnetic moment and magnetic anisotropy in samples on both substrates within their respective switching current ranges. We visualized this thermally assisted magnetization switching process using magnetic force microscopy (MFM), as shown in supplementary material S4, providing annother evidence for the presence of heat during switching. Notably, the rapid heating capability of the mica-based device enables reaching the temperature required to reduce the switching barrier at a significantly lower pulsed current compared to STO. This demonstrates that our mica/SRO/CoPt system effectively reduces the switching current density not only through its intrinsically efficient OT but also synergistically via a heat-assisted mechanism.

For actual flexible applications, mica/SRO/CoPt devices should meet mechanical robustness requirements. Based on our previous work,[48] the mica substrate was thinned to 10 μm and the device adhered to a 55-μm-thick Kapton tape. This configuration provides excellent flexibility (Supplementary Fig. S1), enabling repeated bending. For the test, as illustrated in Fig. 4(a), the device was fixed by a curved mold with a radius of 5mm, and then return to its flat state as one bending cycle. The device was then subjected to $10^3$ such bending cycles. During the cycling process, out-of-plane and in-plane Hall test, as well as magnetization switching loop were measured and shown in Figs. 2(b)-2(d), showing excellent overlap with minimal change. We extracted the anomalous Hall resistance ($R_H$), coercive field ($H_c$), magnetic anisotropy field ($H_k$), and threshold switching current ($J_{th}$) and plotted their evolution in Fig. 4(f). All parameters exhibit negligible change after $10^3$ bending cycles, confirming the robust of both magnetic properties and OT characteristics in our flexible device. Furthermore, electrical robustness test involving 100 consecutive switching loops (Supplementary Fig. S8) confirms the device's excellent endurance under repeated electrical operation.

In summary, we have successfully fabricated and chracterized a flexible OT devices based on a mica/SrRuO$_3$/CoPt heterostructure. The device leverages the prominent OHE in SRO, exhibits a large torque efficiency of -0.31. More importantly, we demonstrate that the exceptional performance of our device stems from a powerful synergy: the intrinsically efficient OT from SRO is magnified by a significant thermally-assisted effect, which arises from the low thermal conductivity of the mica substrate. This synergistic mechanism enables an ultralow switching current density of $9.2\times10^9$ A/m$^2$, a value ~90% lower than that of conventional flexible spin-torque devices and 52% lower than its rigid SRO-based counterpart. Furthermore, the devices exhibits outstanding mechanical robustness, with its performance remaining stable after $10^3$ bending cycles. Our findings establish the flexible OT device as a highly promising paradigm for next-generation flexible spintronics, offering a viable and energy-efficient for wearable and portable applications.

See the supplementary material for details of sample preparation and material characterization, torque efficiency of SRO/CoPt bilayer on STO substure, temperature-dependent resistance and anomalous Hall resistance, pulse-dependent in-plane Hall resistance measurement, magnetic force microscopy image of thermally assisted magnetization switching process, and electrical robustness test.

This work was supported by the National Key Research and Development Program of China (Nos. 2024YFA1410200), the National Natural Science Foundation of China (Nos. 12174406, U24A6001, 52127803), the Chinese Academy of Sciences Project for Young Scientists in Basic Research (No.YSBR-109), the Ningbo Key Scientific and Technological Project (Grant No. 2022Z094).

## AUTHOR DECLARATIONS
**Conflict of Interest**

The authors have no conflicts to disclose.

**Author Contributions**

**Liguang Gong:** Conceptualization (equal); Data curation (equal); Formal analysis (lead); Investigation (lead); Methodology (equal); Visualization (lead); Writing – original draft (lead); Writing – review & editing (equal). **Jian Song:** Conceptualization (equal); Data curation (equal); Formal analysis (equal); Investigation (equal); Methodology (equal); Visualization (equal). **Bin Lao:** Data curation (supporting); Formal analysis (equal); Methodology (equal); Project administration (equal); Software (lead); Writing – review & editing (equal). **Run-Wei Li:** Funding acquisition (equal); Project administration (equal); Resources (equal); Supervision (equal). **Zhiming Wang:** Conceptualization (equal); Data curation (equal); Formal analysis (equal); Funding acquisition (equal); Project administration (equal); Resources (equal); Supervision (equal); Writing – review & editing (equal).

## DATA AVAILABILITY

The data that support the findings of this study are available from the corresponding authors upon reasonable request.

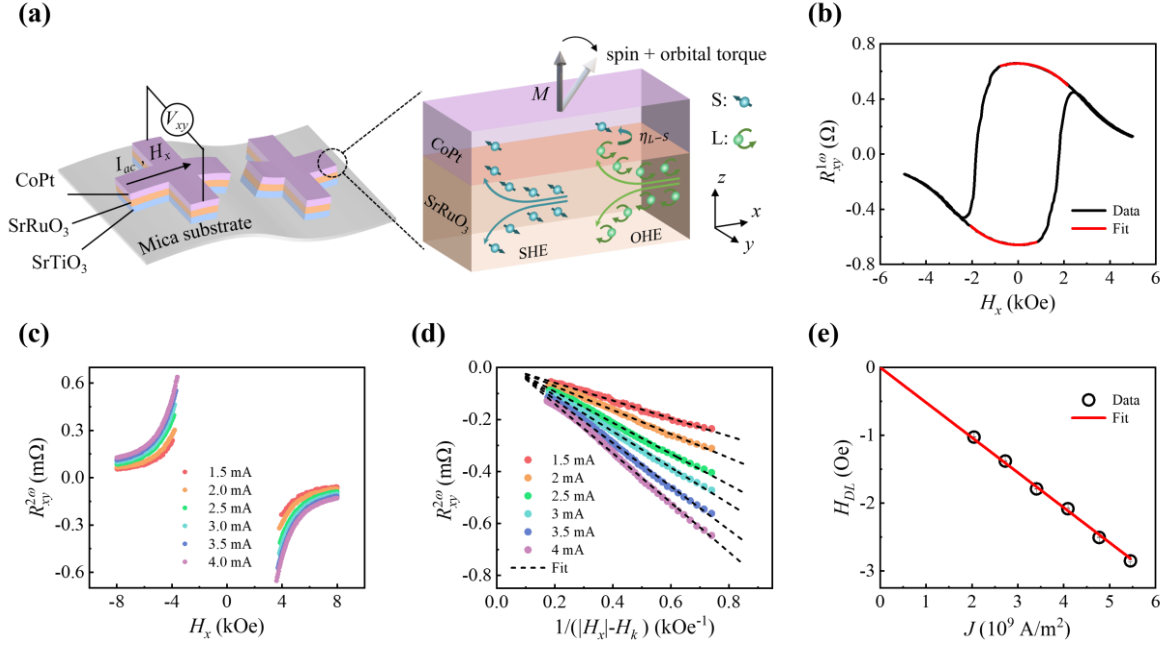

**FIG. 1.** Torque efficiency characterization of the Mica/SRO/CoPt device. (a) Schematic of the harmonic Hall voltage measurement setup. An alternating current ($I_{ac}$) and an in-plane magnetic filed ($H_x$) are applied along the x-direction. The diagram illustrates the generation process of spin torque and orbital torque via SHE and OHE in the SRO/CoPt bilayer, respectively. (b) and (c) are the first harmonic resistance ($R_{xy}^{1\omega}$) and second harmonic resistance ($R_{xy}^{2\omega}$) versus longitudinal swept field ($H_x$), respectively. (d) Second harmonic Hall resistance ($R_{xy}^{2\omega}$) plotted as a function of $1/(|H_x| - H_k)$ with linear fitting. (e) Extracted damping-like effective field ($H_{DL}$) as a function of current density ($J$). The linear fit (red line) to the data yields the torque efficiency.

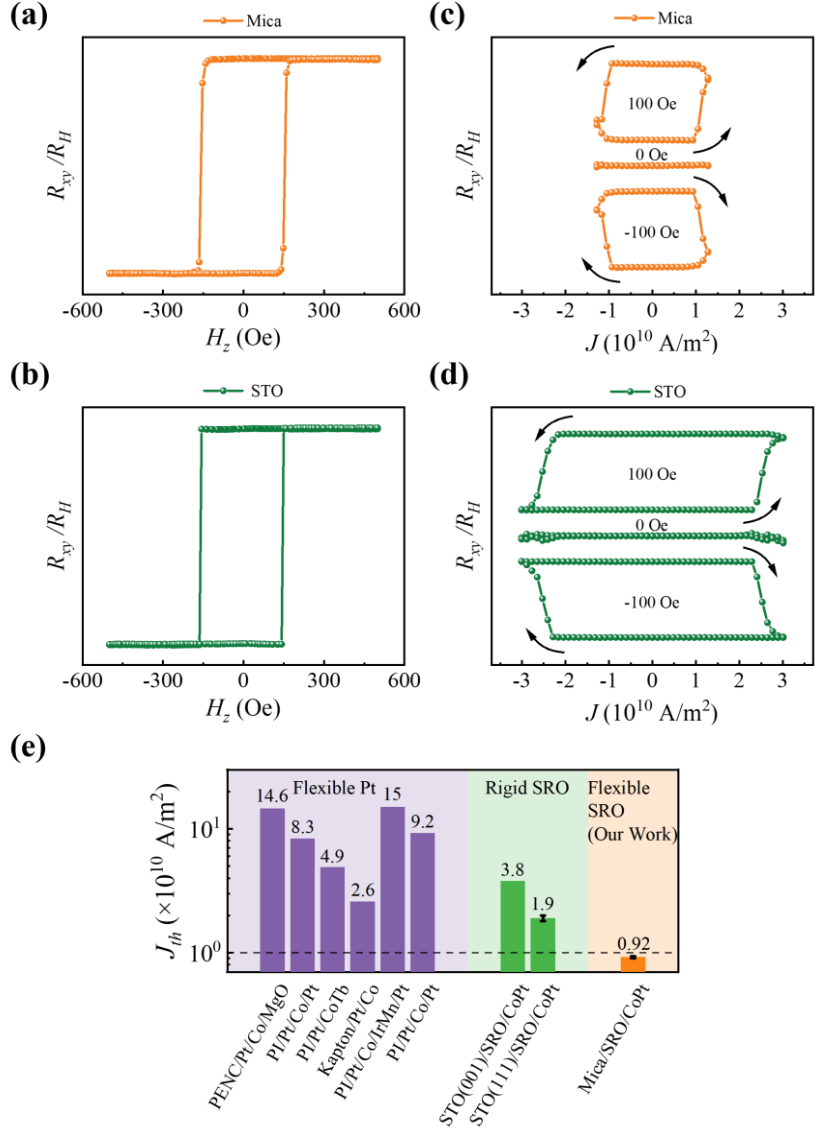

**FIG. 2.** Efficient current-induced magnetization switching in flexible mica/SRO/CoPt devices. (a), (b) $R_{xy}$-$H_z$ hysteresis loops for the devices on mica and the control sample on STO(111), respectively. (c), (d) Current-induced magnetization switching loops under in-plane assistant fields of +100 Oe, 0 Oe, and -100 Oe for the two samples. (e) Comparison of $J_{th}$ values for our flexible/rigid SRO-based devices with state-of-the-art flexible Pt-based spin torque devices and rigid SRO-based OT devices on STO(001). Data for comparison are compiled from Refs. 9, 10, 16-19, and 43.

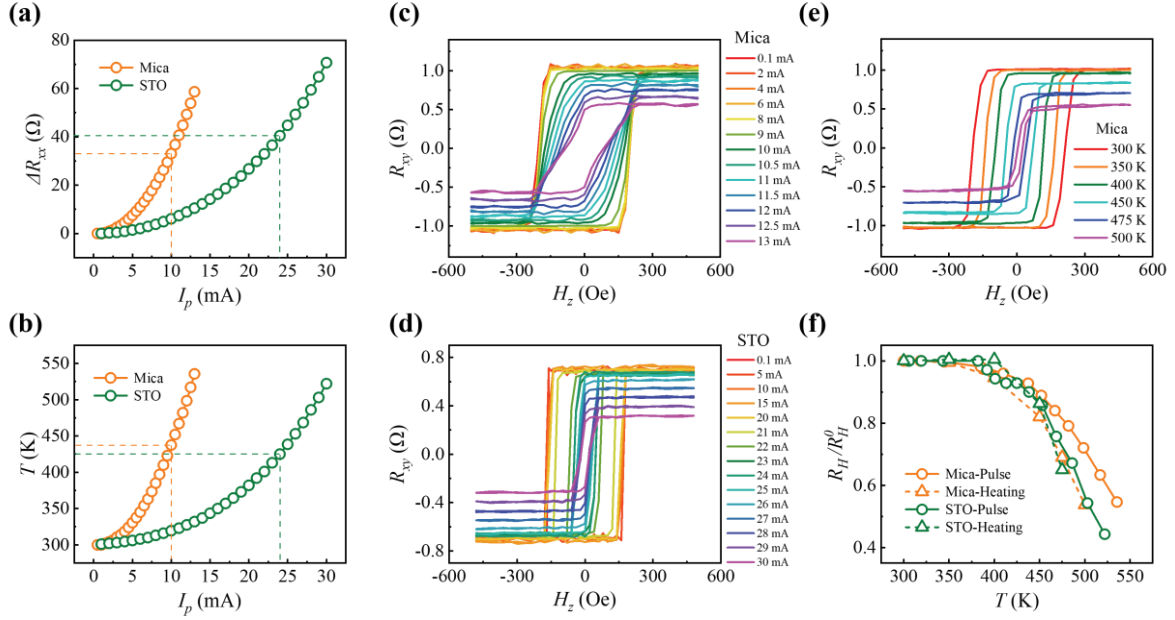

**FIG. 3.** Thermally-assisted switching mechanism in devices on mica and STO(111) substrates. (a) Pulse current-dependent longitudinal resistance change ($\Delta R_{xx}$) for the SRO/CoPt bilayer on mica and STO(111) substrates. The dashed lines indicate the threshold current conditions. (b) Deduced sample temperature variation as a function of pulse current amplitude ($I_p$) for both samples. (c), (d) $R_{xy}$-$H_z$ loops measured under different pulse currents for the mica-based and STO-based devices, respectively. (e) $R_{xy}$-$H_z$ loops measured at different temperatures for the mica-based sample. (f) Unified temperature dependence of normalized anomalous Hall resistance ($R_H / R_H^0$, where $R_H^0$ is defined as $R_H$ measured at 300 K) for both devices under pulse and heating test conditions.

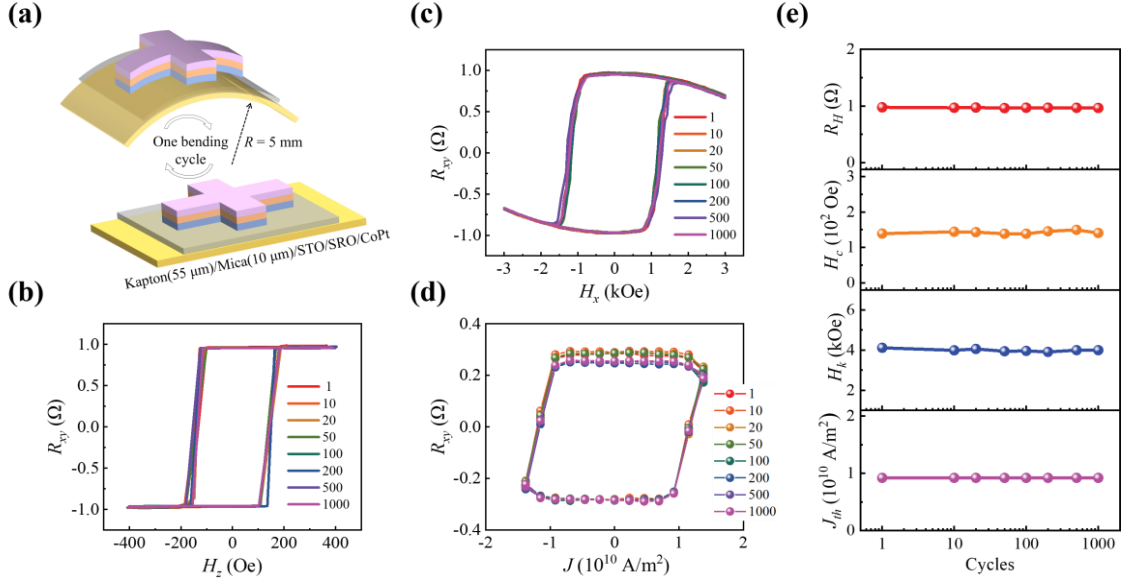

**FIG. 4.** Mechanical robustness of flexible mica/SRO/CoPt OT devices. (a) Schematic of the bending cycle test setup. (b)-(d) Out-of-plane $R_{xy}$ - $H_z$ loop, in-plane $R_{xy}$ - $H_x$ loop, and magnetization switching loops measured after multiple bending cycles. (e) Extracted parameters—anomalous Hall resistance ($R_H$), coercive field ($H_c$), magnetic anisotropy field ($H_k$), and threshold switching current ($J_{th}$)—plotted as functions of bending cycle count.